\documentclass[12pt,preprint]{aastex}
\usepackage{amsmath}
\usepackage{graphicx}
\usepackage{amsfonts}
\usepackage{amssymb}
\shorttitle{}
\shortauthors{Paulo C. Cortes}
\shorttitle{Line Polarization of Molecular Lines}
\shortauthors{P. Cortes}

\begin{document}

\title{Line Polarization of Molecular Lines at Radio Frequencies: The case of DR21(OH)}
\author{P. C. Cortes\altaffilmark{1}
\and R. M. Crutcher\altaffilmark{1}}
\altaffiltext{1}{Astronomy Department, University of Illinois at
Urbana-Champaign, IL 61801}
\and 
\author{W. D. Watson\altaffilmark{2}}
\altaffiltext{2}{Physics Department, University of Illinois at
Urbana-Champaign, IL 61801}

\begin{abstract}
We present polarization observations in DR21(OH) from thermal dust emission at
3 mm and from CO J=$1\rightarrow0$ line emission. The observations were
obtained using the Berkeley-Illinois-Maryland Association array.
\citet{Lai2003} observed this region at 1.3 mm for the polarized continuum
emission, and also measured the CO $J=2\rightarrow1$ polarization. Our
continuum polarization results are consistent with those of \citet{Lai2003}.
However, the direction of the linear polarization for the $J=1\rightarrow0$
is perpendicular to that of the CO $J=2\rightarrow1$ polarization. This
unexpected result was explored by obtaining numerical solutions to the
multilevel, radiative transfer equations for a gas with
anisotropic optical depths. We find that in addition to the anisotropic
optical depths, anisotropic excitation due to a source of radiation that is
external to the CO is needed to understand the orthogonality in the
directions of polarization. The continuum emission by dust grains at the core
of DR21(OH) is sufficient to provide this external radiation. The CO polarization
must arise in relatively low density ( n$_{H_{2}}$ $\sim$ 100 cm$^{-3}$ )
envelope gas. We infer $B \sim 10 \mu G$ in this gas, which implies that
the envelope is subcritical.
\end{abstract}


\section{Introduction}

The star formation process is one of the most complicated problems in current
astrophysical research. Its study includes several physical parameters, of
which the magnetic field is the least observed. Magnetic field observations
are divided into measurements of the Zeeman effect (in order to obtain the
magnetic field strength in the line of sight), and linear polarization
observations of star forming regions. Dust polarization is believed to be
perpendicular to the magnetic field under most conditions \citep
{Lazarian2003}; hence, dust polarization has been used as a major probe for
the magnetic field geometry. In order to efficiently measure the dust
polarization and infer information about the magnetic field morphology, high
resolution observations are required. The BIMA millimeter interferometer has
been used to obtain high-resolution maps in several star forming cores
\citep{Rao1998,Girart1999,Lai2001,
Lai2002,Lai2003}. These results show fairly uniform polarization morphologies
over the main continuum sources, suggesting that magnetic fields are strong,
and therefore, should not be ignored in star formation theory.

The linear polarization of spectral line radiation from the molecular gas has
been suggested to occur under anisotropic conditions \citep{Goldreich1981a}.
The prediction is that  linearly polarized radiation of a few percent is
likely to be detected from molecular clouds and circumstellar envelopes in the
presence of a magnetic field. The direction of this polarization should be
either parallel or perpendicular to the magnetic field, depending on the
angles between the line of sight, the magnetic field, and the direction
associated with the anisotropic excitation \citep{Goldreich1981b}. This
process is known as the Goldreich - Kylafis effect. Combining this effect with
observation of the polarized emission from dust grains, we can probe the
magnetic field morphology in the plane of the sky.

We observed the young and massive star forming region DR21(OH). We measured
dust polarization at 3 mm and the CO $J=1 \rightarrow0$ line polarization.
These observations were compared with the \citet{Lai2003} polarization
observations at 1 mm and CO $J=2 \rightarrow 1$, 
particularly the line polarization that has more extended
emission than the dust. We found a $90^{\circ}$ difference in the position
angle when comparing the line polarization data for the two transitions. 
This difference motivated a
numerical study to explore polarized line emission orientation in different transitions.

The paper is divided into three major sections. In Section 2 we discuss the source
and observation procedure, the line and dust polarization observations, and
the comparison with the \citet{Lai2003} results. The
numerical calculational procedure and results are described in Section 3. 
Section 4 contains the discussion and summary.

\section{Observations and data reduction}

\subsection{Source description}

DR21(OH) is a young massive star forming region located in the -3 km s$^{-1}$
DR21/W75S molecular cloud complex, centered at RA: 20:39:00.7 and DEC:
42:22:46.7 (J2000 coordinates). This region is also at the north-eastern part
of the giant Cygnus-X H II complex. The distance to DR21(OH) is assumed to be
3 kpc; however, the value is uncertain. \citet{Dickel1978} used a value of 2
kpc. From thermal emission of dust \citet{Woody1989} resolved two compact
cores in DR21(OH), MM1 and MM2, with a total mass of $\sim$ 125 M$_{\sun}$.
The MM1 component is the brighter one, with an integrated flux of 0.27 Jy at
2.72 mm \citep{Mangum1991}. DR21(OH) has been extensively mapped in CO by
\citet{Dickel1978,Woody1989,Mangum1991,Chandler1993a}, and in CS by
\citet{Plambeck1990} and \citet{Chandler1993b}. DR21(OH) is also known for its
association with maser emission from OH \citep{Norris1982}, H$_{2}$O
\citep{Genzel1977} and CH$_{3}$OH \citep{Batrla1988,Plambeck1990}. DR21(OH)
has also been observed in the far-infrared \citep{Williams1974,Harvey1986} and
sub-millimeter \citep{Gear1988,Richardson1994}. No centimeter-wavelength
continuum sources have been observed in DR21(OH) \citep{Johnston1984}; 
therefore, HII regions have
not yet developed, so it appears to be in an early stage of evolution. This
makes DR21(OH) a good candidate for early star formation and magnetic field studies.

Magnetic field observations have been carried out measuring Zeeman splittings.
The CN line Zeeman splitting has been detected in both cores (the CN line
traces n$_{H_{2}} \sim$ 10$^{5}$-10$^{6}$ cm$^{-3}$) giving a line-of-sight
magnetic field strength, B$_{los}$ of -0.4 mG for MM1 and -0.7 mG for MM2
\citep{Crutcher1999a}. \citet{Lai2003} observed the CO $J=2 \rightarrow1$
molecular line and the 1.3 mm dust continuum simultaneously using the BIMA
array. They obtained a detailed polarization map for the whole region. Their
results show a remarkably uniform polarization pattern for the line over the
main two continuum sources, while the dust polarization appears to be mostly
perpendicular to the line. These results seem to be consistent with
theoretical predictions.

\subsection{Observation procedure}

We observed DR21(OH) from October to November 2001, mapping the continuum
emission at 3 mm and the CO $J=1 \rightarrow0$ molecular line (at 115.2712
GHz), using the BIMA array in C configuration. We set the digital correlator
in mode 8 to observe both the continuum and the CO $J=1 \rightarrow0$ line
simultaneously. The 750 MHz lower side band was combined with 700 MHz from the
upper side band to map the continuum emission, leaving a 50 MHz window for the
CO line observation (at a resolution of 2.06 km s$^{-1}$). 
Each BIMA telescope has a single receiver and thus the two polarizations
must be observed sequentially.  A quarter wave plate to select either R
or L circular polarization is alternately switched in to the signal path
ahead of the receiver.  Switching between polarizations was sufficiently
rapid (every 11.5 seconds) to give essentially identical uv-coverage.
Cross-correlating the right (R) and left (L) circularly polarized signals from
the sky gave RR, LL, LR, and RL for each interferometer baseline, from which
maps in the four Stokes parameters were produced. The instrumental
polarization was calibrated by observing 3c279, and the ``leakages" solutions
were calculated from these observations. The phase calibrator used was BL Lac.
We used the same calibration procedure described by \citet{Lai2001}.

The Stokes images I, U, Q and V were obtained by Fourier transforming the
visibility data using natural weighting. The 3 mm continuum synthesized beam
had a major axis of 8.4" and a minor axis of 6.9", with a position angle
(P.A.) of 35$^{\circ}$. The MIRIAD \citep{MIRIAD} package was used for data reduction.

\subsection{3 mm Continuum results}

Our 3 mm continuum results did not resolve the two continuum sources (MM1 and
MM2) resolved by \citet{Lai2003}. The larger beam at 3 mm gives poorer
resolution than the 1.3 mm observations (with BIMA at the same configuration).
Our 3 mm continuum observations have a peak emission of 0.22 Jy beam$^{-1}$,
centered at 20$^{h}$39$^{m}$00$^{s}$.7 RA and 42$^{\circ}$22$^{^{\prime}}%
$46$^{"}$.7 DEC in J2000 coordinates. This center coincides roughly with the
center of the MM1 continuum source. \citet{Johnston1984} found no continuum
emission at 14.5 GHz from DR21(OH). Therefore we assume that free-free
emission is negligible and all the continuum radiation comes from dust
emission. The peak emission is consistent with \citet{Mangum1991}, who
obtained 0.192 Jy beam$^{-1}$ at 2.7 mm with a comparable beam size.

Polarization detection is very sensitive to the signal to noise ratio. Because
of bad weather conditions during part of our observations we did not achieve
the same level of polarization sensitivity as \citet{Lai2003} did. We have
3$\sigma$ polarization detections (figure 1) scattered over 3 positions in the
map. The dust polarization detected is consistent with the result of
\citet{Lai2003}. 

\subsection{CO $J=1 \rightarrow 0$ Observational results}

Figure 2 shows the Stokes I spectra integrated over a region that contains the
MM1 and MM2 sources \citep{Lai2003}. This region covers the box from RA:
20$^{h}$39$^{m}$2$^{s}$ - 20$^{h}$38$^{m}$59$^{s}$.5 to DEC:42$^{\circ}$22'35"
- 42$^{\circ}$23'2". Plotted over the Stokes I 
are the P.A. and the fractional polarization
at each velocity channel. The Stokes I spectra generally agree with the CO
$J=2 \rightarrow 1$ spectra from \citet{Lai2003}. In our case the spectrum has
a peak at $v_{lsr}=-8$ km s$^{-1}$, a minimum at $v_{lsr}=-2 $ km s$^{-1}$, and
a second peak at $v_{lsr}=2$ km s$^{-1}$. 
The velocity of these peaks are different from previous CS and CN observations,
 which are at -5 km s$^{-1}$ \citep{Chandler1993b,Richardson1994}, 
and -5 km s$^{-1}$ and -1 km s$^{-1}$ \citep{Crutcher1999a}. 
The CO line is optically thick (Figure 2) and the dip, or zero emission range,  
observed in both CO $J=2 \rightarrow 1$ and $J=1 \rightarrow 0$ lines, 
is probably due to self-absorption. 
In CO $J=1 \rightarrow 0$, the dip covers a velocity range from -6 km s$^{-1}$ to 
2 km s$^{-1}$ which coincides with the 
C$^{34}$S $J=2 \rightarrow 1$ emission peak range from \citet{Chandler1993b}, 
which goes from -6 km s$^{-1}$ to 1.5 km s$^{-1}$. In the same velocity range
$^{13}$CO emission was detected \citep{Dickel1978}  from
 -3 km s$^{-1}$ to 0.5 km s$^{-1}$. C$^{34}$S traces higher densities
than CO, and $^{13}$CO is more optically thin than CO. The presence of
peak emission from these lines in the same velocity range of our dip 
suggests that our CO observations may come from a different
region from the higher density gas (which probably comes from the core). 
Additionally, it is expected
that CO polarized line emission will arise from optically thin regions
($\tau \sim 1$) \citep{Goldreich1981a,Watson1984}.  
Therefore, it is likely
that the CO polarized emission that we detected comes from an envelope
around DR21(OH).

Figure 3 shows our CO polarization map at $v_{lsr}=$-8 km s$^{-1}$. This map
represents the highest significance polarization distribution in our
observations. It also coincides with the peak emission of the CO. Figure 4
shows our polarization map at $v_{lsr}=$-10 km s$^{-1}$, which is the same
velocity map presented by \citet{Lai2003}. 
In our case the $v_{lsr}=$-10 km s$^{-1}$ map
does not present the same spatial distribution of polarization as the -8 km
s$^{-1}$ map, almost certainly due to the limited sensitivity of the
polarization data. Table 1 gives a comparison between P.A. for both CO
transitions. We can see that there is a consistent 90$^{\circ}$ difference
between the transitions.

\subsection{Comparison between CO $J=1 \rightarrow0$ and CO $J=2 \rightarrow1$ polarization}

\citet{Lai2003} presented a single velocity channel map of their line
polarization observations (at $v_{lsr}=-10$ km s$^{-1}$); this is the velocity of 
the  peak value in the Stokes I emission.  
Our CO $J=1 \rightarrow0$
observations have a peak Stokes I at $v_{lsr}=-8$ km s$^{-1}$ (Figure 3
), which also coincides with our most complete polarization spatial
distribution. 

The \citet{Lai2003} map shows distributed CO emission that is more extensive
than  the region of the 1.3 mm
continuum sources. We see a similar situation in our observations (Figure 3). We
also see that our line polarization has better coverage over the CO emission
than the \citet{Lai2003} observations (Figure 4). Our observations seem to
indicate that with better signal to noise, it might be possible to achieve
complete polarization coverage over the region of CO emission. The 
maximum polarized intensity coincides with the position of the continuum
source, a fact which  suggests the importance of a continuum radiation field in polarizing
the line emission. This was studied in our numerical calculation.

Comparing polarization vectors from Figure 3 and Figure 4 with the map
presented by \citet{Lai2003}, we found that the P.A.
differ; in the central region in both maps most of the P.A.
are orthogonal. This is a rather unexpected discovery. The Goldreich - Kylafis
effect predicts that polarization can be either parallel or perpendicular to
the magnetic field \citep{Goldreich1981a}, but did not make a distinction
between polarization of the same molecule at different transitions. This is
because their calculation was done simulating a 2-level molecule.
\citet{Watson1984} extended the calculation to a multilevel molecule, but made
no prediction about polarization direction in different transitions. Direct
comparison of angles in Table 1 shows the $90^{\circ}$ difference between both
CO transitions. This is particularly true in the central part of the map. 

\section{Polarization of molecular lines}

\subsection{The Goldreich - Kylafis effect}

The Goldreich - Kylafis effect is described in a series of papers
\citep{Goldreich1981a,Goldreich1981b,Kylafis1983a,Kylafis1983b}. They
considered a molecule with only two (rotational) states 
having angular momenta $1$ and $0$, and investigated line formation
and polarization in the presence of a magnetic field and anisotropic optical
depths. Their original work predicted linear polarization up to about ten per
cent. 
Motivated
in part by an unsuccessful observational survey for this polarization
\citep{Wannier1983}, \cite{Watson1984} extended the calculations to include a
number of rotational states and radiative transitions. As a result, the
predicted linear polarizations were reduced, typically by a factor of about
two. We have further extended the multilevel calculations by including an external source
term to represent emission from dust in a compact source. We have also used
improved rates for collisional excitation, though this proved to be
unimportant. We focus on calculations to understand the polarization angles of
the CO $J=2\rightarrow1$ and $J=1\rightarrow0$ transitions observed in DR21(OH).

\subsection{Basic methods}

The formulation used here follows \cite{Watson1984}. The radiative
transfer equations are solved in the large velocity gradient or Sobolev
approximation (hereafter LVG), and in the regime where the line width
$\Delta\nu$, the Zeeman splitting $g\mu_{0}B$, and the natural 
line width $\gamma$ obey the relation $\Delta\nu\gg g\mu_{0}B\gg\gamma$. This strong
inequality is easily satisfied here. The quantization $z$ axis is along the
magnetic field direction, and the quantum states are specified in the usual
way by the total angular momentum $J$ (there is no fine or hyperfine structure here)
and by its projection $M$ on the $z$ axis. Under these conditions
the radiative transfer equations for radiation associated with transitions 
between  upper state $J$ and lower state $J^{\prime}$ can be written as%

\begin{align}
\frac{dI_{JJ^{\prime}}^{\perp}}{ds}=-\kappa_{JJ^{\prime}}^{\perp
}(I_{JJ^{\prime}}^{\perp} - S_{JJ^{\prime}}^{\perp})\\
\frac{dI_{JJ^{\prime}}^{\parallel}}{ds}=-\kappa_{JJ^{\prime}}^{\parallel
}(I_{JJ^{\prime}}^{\parallel} - S_{JJ^{\prime}}^{\parallel})
\end{align}

\noindent where $\parallel(\perp)$ indicates the intensity of the radiation
with a linear polarization parallel\ (perpendicular) to the magnetic field.

The difference with \cite{Goldreich1981a} comes in expressing the opacity and
source terms in a way to allow for arbitrary angular momenta. The opacity and
source terms are thus written as
$\kappa_{JMJ^{\prime}M^{\prime}}$ and $S_{JMJ^{\prime}M^{\prime}}$ such that
$\kappa_{JJ^{\prime}}^{q}$ and $S_{JJ^{\prime}}^{q}$ (where $q$ stands for
$\parallel$ or $\perp$), 

{\small
\begin{equation}
\kappa_{JJ^{\prime}}^{\perp}=\frac{1}{2}\phi(\nu- \nu_{JJ^{\prime}}) \sum_{\Delta M=1}
\kappa_{JMJ^{\prime}M^{\prime}}
\end{equation}%
\begin{equation}
S_{JJ^{\prime}}^{\perp}=\frac{\sum_{\Delta M=1} \kappa_{JMJ^{\prime}M^{\prime
}} S_{JMJ^{\prime}M^{\prime}}} {\sum_{\Delta M=1} \kappa_{JMJ^{\prime
}M^{\prime}}}
\end{equation}%
\begin{equation}
\label{absorption1}
\kappa_{JJ^{\prime}}^{\parallel}=\phi(\nu- \nu_{JJ^{\prime}}) \left(
sin^{2}\theta\sum_{\Delta M=0} \kappa_{JMJ^{\prime}M^{\prime}}+\frac{1}{2}
cos^{2}\theta\sum_{\Delta M=1} \kappa_{JMJ^{\prime}M^{\prime}} \right) 
\end{equation}%
\begin{equation}
\label{emission1}
S_{JJ^{\prime}}^{\parallel}=\frac{sin^{2}\theta\sum_{\Delta M=0}
\kappa_{JMJ^{\prime}M^{\prime}} S_{JMJ^{\prime}M^{\prime}} + \frac{1}{2}
cos^{2}\theta\sum_{\Delta M=1} \kappa_{JMJ^{\prime}M^{\prime}}S_{JMJ^{\prime
}M^{\prime}}} {sin^{2}\theta\sum_{\Delta M=0} \kappa_{JMJ^{\prime}M^{\prime}}
+ \frac{1}{2} cos^{2}(\theta)\sum_{\Delta M=1} \kappa_{JMJ^{\prime}M^{\prime}%
}}%
\end{equation}
}

{\normalsize \noindent where the symbol $\theta$ represents the angle between
the magnetic field and the direction of propagation and $\kappa_{JMJ^{\prime
}M^{\prime}}$ and $S_{JMJ^{\prime}M^{\prime}}$ are defined in terms of
Einstein A-coefficients and the populations $n_{JM}$ per magnetic substate}

{\normalsize
\begin{align}
\kappa_{JMJ^{\prime}M^{\prime}}  &  =\frac{3}{8\pi}\left(  \frac{c}%
{\nu_{JJ^{\prime}}}\right)  ^{2}A_{JMJ^{\prime}M^{\prime}}(n_{J^{\prime
}M^{\prime}}-n_{JM})\label{absorption3}\\
S_{JMJ^{\prime}M^{\prime}}  &  =\frac{h\nu_{JJ^{\prime}}^{3}}{c^{2}}%
\frac{n_{JM}}{n_{J^{\prime}M^{\prime}}-n_{JM}}.
\end{align}
\newline Some equations here differ slightly from those in \cite{Watson1984}
because we are treating all magnetic substates explicitly. All statistical
weight factors are one and are thus omitted for simplicity.}\newline 

{\normalsize Since the radiative transfer equations are functions of the
populations, rate equations must be solved for these populations. We assume
steady state , $\frac{\partial n_{i}}{\partial t}=0$. The rate equations for
the populations per magnetic substate are }

{\normalsize
\begin{equation}
\frac{\partial n_{i}}{\partial t}=-\sum_{j<i}P_{ij} + \sum_{j>i}P_{ji} + \sum_{j}(C_{ji}n_{j}
- C_{ij}n_{i}) = 0
\end{equation}
}

{\normalsize \noindent where the $P_{ij}$ \ (which involve the radiation) are
given in terms of the Einstein $A_{JMJ^{\prime}M^{\prime}}$ coefficients and
the stimulated emission coefficients $R_{ij}$. The indices $i,j$ correspond to
$(J,M),(J^{\prime},M^{\prime})$. The $C_{ij}$ are the collisional excitation
rates from state }$i$ to state {\normalsize $j$}.
{\normalsize \cite{Watson1984} used the $C_{ij}$ given by \cite{Green1978}. In
our calculation we have updated the values of $C_{ij}$ with 
the values provided by \cite{Flower2001}, which are given for 
a wide range of gas temperatures. We used 
a weighted average for the contribution of ortho and para hydrogen 
cross sections, specifically,  $C_{ij}=0.7\times C_{ij}^{para}+0.3\times
C_{ij}^{ortho}$.  
We considered temperatures up to $T_{gas}=200K$ and angular momenta up to $J=9$.
The expressions for the $P_{ij}$ are }

{\normalsize
\begin{equation}
\label{MacroPol}
P_{ij}=A_{ij}[n_{i}+R_{ij}(-n_{j}+n_{i})].
\end{equation}
The population equations constitute a nonlinear system, which we solve by
numerical iteration. }

{\normalsize In the limit in which the macroscopic velocity differences of the
gas in the cloud are greater than the thermal velocities of the molecules, the
LVG approximation can be used to simplify the solution for the radiative
transfer equations. The LVG approximation expresses the intensity as a
function of local variables. The intensity that emerges from the gas cloud and
is detected by an observer is}

{\normalsize
\begin{equation}
I_{JJ^{\prime}}^{q}=\left(  S_{JJ^{\prime}}^{q}-\frac{B}{2}\right)
[1-exp(-\tau_{JJ^{\prime}}^{q})]\label{LVG}%
\end{equation}
where }$B$ represents the cosmic background radiation which is included in the
calculation, and $\tau_{JJ^{\prime}%
}^{q}$ is the LVG optical depth for radiation with polarization $q$ at the
frequency of the $JJ^{^{\prime}}$ transition,

\begin{equation}
\tau_{JJ^{\prime}}^{q}=\kappa_{JJ^{\prime}}^{q}\times L(\theta)
\end{equation}
and where $L(\theta)$ is the LVG characteristic scale length for which we
adopt the form,

\begin{equation}
L(\theta)=L_{0}/(\alpha\sin^{2}(\theta)+\cos^{2}(\theta)).
\end{equation}

When $\alpha<<1,L(\theta)$ reflects a velocity gradient that is perpendicular
to the magnetic field; for $\alpha>>1$, the velocity gradient is along the magnetic field
(e.g., \citet{Watson1984}). In the calculation, the constant
$L_{0}$ enters multiplied by the total number density of CO molecules. We vary
this product to obtain the solutions as a function of optical depth that are
presented in the Figures.

The $R_{ij}$ in equation \ref{MacroPol} involve the integral over frequency and over solid angle
 of the intensity within the cloud at the location of the CO, and are
{\normalsize \ non-zero only for $|J-J^{\prime}|=1$, $|M-M^{\prime}|=0$ and
$|M-M^{\prime}|=1$ \citep{Watson1984},}

{\normalsize
\begin{equation}
\label{Rrates1}
R_{JJ^{\prime}}^{0}=\frac{3c^{2}}{2h\nu^{3}}\int\frac{d\Omega}{4\pi}%
sin^{2}\theta\int d\nu\phi(\nu-\nu_{JJ^{\prime}})I_{JJ^{\prime}}^{\parallel
}(\Omega)
\end{equation}
}

{\normalsize
\begin{equation}
\label{Rrates2}
R_{JJ^{\prime}}^{1}=\frac{3c^{2}}{4h\nu^{3}}\int\frac{d\Omega}{4\pi}\int
d\nu\phi(\nu-\nu_{JJ^{\prime}})[I_{JJ^{\prime}}^{\perp}(\Omega)+cos^{2}\theta
I_{JJ^{\prime}}^{\parallel}(\Omega)].
\end{equation}
}In the LVG approximation the integrals over frequency reduce to

{\normalsize
\begin{equation}
\int I_{JJ^{\prime}}^{q} \phi(\nu- \nu_{JJ^{\prime}})d\nu= S_{JJ^{\prime}}%
^{q}(1 - \beta_{JJ^{\prime}}^{q}) + \frac{B+S(\Omega)}{2} \beta_{JJ^{\prime}}^{q}%
\end{equation}
}

{\normalsize \noindent in which the escape probability function is
$\beta_{JJ^{\prime}}^{q}=[1-exp(-\tau_{JJ^{\prime}}^{q})]/\tau_{JJ^{\prime}%
}^{q}$, and S($\Omega$)  represents the radiation that is incident from a compact,
external continuum source. We express }S$(\Omega)$ as

\begin{equation}
\label{compactsource}
S(\Omega)=(1-e^{-\tau_{c}})B_{\nu}(T_{source})
\end{equation}

\noindent for directions $\Omega$ that are subtended by the external source as
viewed from the location of the CO. In other directions, S$(\Omega)=0$. In equation
\ref{compactsource}, B$_{\nu}(T_{source})$ is the Planck function,
$T_{source}$ is the source temperature, and $\tau_{c}$ is the continuum
optical depth of the source at the relevant frequency. We adopted the parameterization
of \citet{Mangum1991} for the combined emission from the dust core of DR21(OH), 
specifically, T$_{source}=42$ K, $\tau_{c}$=1.5$[0.25/ \lambda [mm]]^{2}$ 
and an effective source angular
size (radius) of 10$^{\prime \prime}$ as seen from the earth. At the location of the CO, 
the angular size (radius)
for the source is then approximately 1/2 radian (assuming 0.3 pc 
for the separation between the CO and the source).
{\normalsize Finally, the fractional polarization is expressed as }

\begin{equation}
P=\frac{I_{\perp}-I_{\parallel}}{I_{\perp}+I_{\parallel}} \label{fpol}.
\end{equation}

\subsection{{\protect\normalsize Calculation results}}

{\normalsize The objective of this calculation is to understand why the the
directions of the polarizations for the CO $J=1\rightarrow0$ and the CO
$J=2\rightarrow1$ radiation are perpendicular to one another in the DR21(OH)
observations described in Section 2, with the $J=1\rightarrow0$ polarization
parallel to that of the emission by the dust grains. We want to calculate the
linearly polarized radiation emitted by CO for physical conditions that are
representative for the DR21(OH) star forming region, and determine whether the
observed directions and magnitudes of the 
polarization can be reproduced. The orthogonality of the
polarizations is reflected by a sign difference in }$P$ in equation
{\normalsize \ref{fpol}} {\normalsize. If $P>0$, $I_{\perp} $ dominates the
emission and the polarization is perpendicular to the magnetic field. In the
other case it is parallel. }

{\normalsize The physical conditions for our multilevel calculation are
described below. These conditions are consistent with the assumption that the polarized
emission comes from a cold envelope around the DR21(OH) continuum sources. We
assumed that the physical conditions are the same in the gas for both CO
transitions. }

To obtain a qualitative understanding of how the polarizations can be
orthogonal for the two transitions, consider the $J=1\rightarrow0$
transition by itself, and first without any external radiation. \ If the
velocity gradients are smaller in directions parallel to the magnetic field
than in directions perpendicular to the magnetic field, \ the optical depths
for spectral lines will be the largest parallel to the field lines. \ The
escape of radiation involved in de-exciting the upper $J=1$ state will then be
reduced more in directions along the field lines than in directions
perpendicular to the field lines. This in turn leads to populations of the
$M=\pm1$ substates that are larger than the populations of the $M=0$ substate
because of the difference in the angular distributions of the de-exciting
radiation. The angular distribution of the $\sigma$ radiation ($\left|  M-M%
\acute{}%
\right|  $ $=1$ transitions) peaks in directions along the field lines whereas
that of the $\pi$ radiation ($\left|  M-M%
\acute{}%
\right|  $ $=0$ transitions) peaks in directions perpendicular to the magnetic
field. Hence, when the velocity gradients are smallest along the
magnetic field, the rate for de-excitation of the $M=\pm1$ substates is
decreased more by trapping of the radiation than is the de-excitation rate of
the $M=0$ substate . The excitation rate for all magnetic substates by
collisions is the same. Under isotropic conditions, the populations of the
magnetic substates are equal. For any direction, the $\sigma$ and $\pi$
radiation that emerges from the radiative decays of these states will combine
to give zero net polarization. However, when the $M=\pm1$ states have larger
populations, the $\sigma$ radiation will be relatively stronger in comparison
with the $\pi$ radiation. Their contributions to the polarization will not
then cancel and the combined radiation will be linearly polarized---in the
direction of the polarization of the $\sigma$ radiation which is perpendicular
to the magnetic field.

Now consider by itself a compact, external source of radiation 
(i. e., no anisotropy in the optical depths as in the above paragraph)
that is located in a
direction perpendicular to the magnetic field. Because the angular
distribution associated with $\pi$ transitions is larger for this direction
than is the angular distribution of the $\sigma$ transitions, the $J=1,M=0$
substate will be excited more rapidly by absorption of this radiation than
will be the $J=1,M=\pm1$ substates. The
$J=1,M=0$ substate would then become overpopulated in comparison with the
other two magnetic substates and radiation that emerges from the gas would be
polarized in the direction of $\pi$ radiation---parallel to the magnetic field.

Though the reasoning is more complicated because of the additional substates
and transitions, analogous conclusions apply for the polarization of the
radiation associated with the $J=2\rightarrow1$ and higher transitions of the
CO molecule.

When the above anisotropic velocity gradients and external source of radiation
are both present, the direction of the polarization of the radiation that is
emitted by the gas is determined by the relative importance of these two
causes of anisotropy. Their relative importance is not, however, the same for
the $J=2\rightarrow1$ as for the $J=1\rightarrow0$ transition. The dust radiation
from the core of DR21(OH) is much higher at the CO $J=2 \rightarrow 1$ transition
 frequency than the CO $J=1 \rightarrow 0$ frequency, and thus the external 
radiation will play a greater role for the
polarization of the $J=2\rightarrow1$ transition than it does for the
$J=1\rightarrow0$ transition. The effect of the anisotropic velocity gradients
is about the same for the two transitions since the collisional excitation rates
are about the same for both transitions
(the excitation energies for
these states of CO are less than $kT_{gas}$). The $J=2\rightarrow1$ radiation
can then reflect the influence of the radiation and 
be polarized parallel to the magnetic field under conditions for
which the $J=1\rightarrow0$ radiation reflects the influence of the
velocity gradient and is polarized perpendicular to the
magnetic field. The results presented below from the numerical calculations
demonstrate that this orthogonality in the polarizations occurs for physical
conditions that are likely for the gas that is emitting the polarized CO
radiation from DR21(OH). Velocity gradients that are anisotropic and
systematically larger in the direction perpendicular to the magnetic field
occur in magnetohydrodynamic turbulence (see \citet{Watson2004})

Calculations were performed for gas temperatures between
15 and 50 K and for a range of densities. 
\citet{Dickel1978} obtained a T$_{gas}$=28 K;
this temperature is used as a reference for our calculations. 
The velocity gradient must be largest in the direction
perpendicular to the magnetic field according to the above reasoning. We
thus focus on $\alpha=0.1$ in the expression for $L(\theta)$, though we also
present results for $\alpha$=0.3 for comparison.

{\normalsize Three curves for polarization versus LVG optical depth 
that correspond to calculations with densities
$n_{H_{2}}=$ 25, 75 and 225 cm$^{-3}$ are shown in Figure 5 
for $T_{gas}=28$ K. 
These densities are in agreement with what we might expect in a lower
density envelope. \citet{Padin1989} obtained a $n_{H_{2}}$ density of
$n_{H_{2}}=5\times10^{6}$ cm$^{-3}$, but this corresponds to the MM1 continuum
source and not to the DR21(OH) envelope. 
At the lowest optical depths, the
external continuum source dominates in determining the direction of the
polarization for both transitions whereas at the highest optical depths this
radiation is largely excluded and the velocity gradient determines the direction of
the polarization for both transitions. However, because  the continuum emission
from the core of DR21(OH) is much higher at the frequency of the $J=2 \rightarrow 1$
transition than at the frequency of the $J=1 \rightarrow 0$, the external radiation
remains more important to higher optical
depths for the $J=2\rightarrow1$ transition than for the $J=1\rightarrow0.$
Hence, a range of optical depths occurs over which the polarizations are
orthogonal. This range of optical depths is mostly from }$\log(\tau
)\simeq 0$ to $-1$ in Figure 5. In this range of optical depths,
the fractional polarizations typically
are a few percent and are thus consistent with the observed magnitudes for the
fractional polarizations
of the two transitions. {\normalsize  These small $\tau$ values are consistent
with our assumption that the polarized CO emission arises from a cold, low
density gas, which in turn is compatible with the likely conditions for the
DR21(OH) envelope. At the highest optical depths, the fractional polarization
tends to zero in Figure 5 for both transitions as expected.} The polarizations
in the second panel in Figure 5 (for $\alpha$=0.3) provide an indication of the
sensitivity of the calculations to the magnitude of the velocity gradients.
The polarizations for both transitions are presented only as a function of the
CO $J=1 \rightarrow 0$ optical depth. However, the CO $J=2 \rightarrow 1$
optical depth is similar in our calculations.

The curves in
Figure 5 correspond to the densities that represent our best results. Our
calculations show that for densities over n$_{H_{2}}$=500 cm$^{-3}$
the fractional polarization decreases dramatically for both lines,
vanishing almost completely for densities over n$_{H_{2}}$=1000 cm$^{-3}$.
At the higher densities collisions are more rapid and the continuum source 
is no longer an important source of anisotropy.
On the other hand, densities below n$_{H_{2}}$=20 cm$^{-3}$ would require
an envelope too big and far away from the continuum source, which would make 
the central continuum source 
ineffective as an additional source of anisotropy.

{\normalsize  In Figure 6, we demonstrate explicitly that the polarization is
positive for both transitions at all optical depths in our calculations 
when there is no continuum source so that  the velocity gradients
determine the direction of the polarization. The fractional polarization also
tends to zero at large and at small optical depths as expected, and as found by
the previous investigators.}

In {\normalsize Figure 7, we present the results of our calculations for the
other extreme; the limit in which the radiation from the continuum source
dominates. We cannot eliminate collisions completely in our calculations.
However, we can greatly reduce their influence by multiplying the collision
rates by a small factor (10}$^{-5}$). {\normalsize  As expected, the
polarization is now negative for both lines (parallel to the magnetic field),
is independent of density
and converges to zero for higher optical depths. 

{\normalsize In our calculations, we also varied the gas temperature in order
to explore the influence of uncertainty about the gas temperature in the
envelope around DR21(OH). We found no significant variation in the polarizations
for $T_{gas}$ between 15 and 50 K. What we found is that the
range in }$\tau$ for which the polarizations are orthogonal {\normalsize
shifts slightly in $\tau$; there is a displacement to lower $\tau$ for higher
temperatures. A molecular envelope might be heated by external UV photons or
cosmic rays. This insensitivity to gas temperature shows a degree of
robustness for the results of our calculations.} We have also verified that 
the results are insensitive to the temperature of the dust by performing 
computations with $T_{source}$ up to 60 K.

{\normalsize The polarizations presented in Figures 5-7 are for the
radiation that is emitted by the gas at an angle of 90$^{\circ}$ with respect to the
magnetic field. We have verified that the polarization characteristics are
similar for emission at angles from 90$^{\circ}$ to about 40$^{\circ}$. At
smaller angles, the polarized fraction decreases significantly---approaching
zero at 0$^{\circ}$,} as expected.

\subsection{{\protect\normalsize Magnetic field strength and cloud support}}

{\normalsize Our work has shown that the line polarization will trace magnetic
field geometry at lower densities than the dust polarization. However,
comparing with the results of \citet{Lai2003}, we believe that there is a
correlation between the field traced by the dust (core) and the field traced
by the line (envelope). Our dust polarization result at 3 mm shows similar
P.A. to the 1.3 mm dust polarization of \citet{Lai2003}. Also we now know why
the line polarizations are perpendicular at two different transitions, which
makes the CO $J=1 \rightarrow0$ polarization consistent with the CO $J=2
\rightarrow1$. At different frequencies the line polarization is either
parallel or perpendicular to the dust polarization. 
The agreement in the geometries of the 
dust and line polarization suggests that the field
in the core and the envelope are connected and not independent. }

{\normalsize \citet{Lai2003} estimated the magnetic field strength to be about
0.9 mG for MM1 (n$_{H_{2}} \sim 1 \times 10^{6}$ cm$^{-3}$) and 1.3 mG for MM2
(n$_{H_{2}} \sim 2 \times 10^{6}$ cm$^{-3}$). 
\cite{Crutcher1999a} measured a magnetic
field in the line of sight $B_{los} \sim$ -0.4 mG for MM1 and $B_{los} \sim$
-0.7 mG for MM2. \citet{Chandrasekhar1953} predicted a magnetic field dependence
proportional to $\rho^{1/2}$ }

{\normalsize
\begin{equation}
B_{ChF}=Q\sqrt{4\pi\rho} \frac{\Delta v_{los}}{\Delta\phi}%
\end{equation}
}

The dispersions in the measured velocities  and P.A. are similar 
for the $J=1 \rightarrow 0$ and $J=2 \rightarrow 1$ lines. Therefore,
we can extrapolate the values of the magnetic field
measured in the core to the envelope following a $\rho^{1/2}$ law. The
molecular hydrogen number density for the cores is in the order of n$_{H_{2}}
$ = 10$^{6}$ cm$^{-3}$, and taking a density of n$_{H_{2}}$ = 100 cm$^{-3}$
for the envelope, we get a magnetic field in the envelope 100 times smaller
than the core value of $\sim$ 1 mG (at n$_{H_{2}} \sim 1 \times 10^{6}$ cm$^{-3}$), 
or about 10 $\mu G$. We can use this value to estimate
the mass to magnetic flux ratio; its critical value is given by
\citep{Mouschovias1976} }

{\normalsize
\begin{equation}
\label{massTOflux}
\frac{M}{\Phi_{B}}_{crit} = \frac{1}{\sqrt{63 G}}%
\end{equation}
}

{\normalsize \noindent This equation can be expressed as a function of volume
density in an spherical model, using the molecular hydrogen mass, equation
\ref{massTOflux}, and expressing the magnetic field in G units, and the
distance in cm. We arrived at }

{\normalsize
\begin{equation}
\frac{M/ \Phi}{M/ \Phi}_{crit}=9.1 \times10^{-27} \frac{R [cm] \times n
[cm^{-3}]}{B[G]}%
\end{equation}
}

{\normalsize \noindent in which R is the radius of the cloud, $n$ is the number
density of hydrogen molecules and B is the magnetic field. Using a density
$n=$100 cm$^{-3}$, $B$=10 $\mu$G and a radius of R=0.3 pc, we obtained a
mass-to-flux ratio of 0.13 times the critical value,
 and thus highly subcritical. Hence, the envelope
is supported against gravitational contraction by the magnetic field.
The value of R represents an
angular distance of 0.3 pc of distance for DR21(OH). This radius may be
underestimated due to resolution of more extended structure by the
interferometer. However, single dish maps \citep{Wilson1990} suggest that R
$<$ 1 pc. }

\section{{\protect\normalsize Summary and conclusions}}

{\normalsize DR21(OH) was observed in 3 mm dust continuum and CO $J=1
\rightarrow0$ line emission. Comparing our line observations with
\citet{Lai2003}, we observed a consistent difference in polarization direction
between the CO $J=2 \rightarrow 1$ and CO $J=1 \rightarrow 0$ lines; 
they are perpendicular to each other over the central region
(which coincides with the position of the MM1 and MM2 continuum sources). We
developed a code based on the \cite{Watson1984} calculation in order to solve
the radiative transfer equations and calculate the fractional linear
polarization for different transitions of the CO molecule. We found that the
presence of a small continuum source will likely produce an increase in the
anisotropy of the radiation field over the CO gas. An anisotropic radiation
field will unevenly populate the magnetic sub-levels of the CO molecule, which
will produce linearly polarized emission. We showed that the presence of a
compact continuum source will produce a gradual change in the fractional polarization,
with a sign change, as a function of optical depth. This would explain the orthogonal
orientations of the CO polarization in different transitions. In the
particular case of DR21(OH), the physical conditions that are consistent with
the polarization data correspond to a hot continuum source ($T_{dust}$ $\sim$ 42 K)
and a CO gas temperature of $T_{gas} \sim$ 28 K, for a $n_{H_{2}} \sim $100
cm$^{-3}$. }

{\normalsize Line polarization observations can become a powerful tool to
constrain the magnetic field geometry. We have seen that a clear understanding
of the physical conditions in a molecular cloud will give more accurate
information about the fractional polarization sign. A good knowledge of the
sources of anisotropy in a cloud will help to understand and constraint the
geometry in line polarization observations at multiple frequencies. }

{\normalsize We also found that line polarization traces low density material;
this is important in order to connect the magnetic field geometries at
different densities. In the particular case of DR21(OH) we believe that there
may be a correlation between the field traced by the dust at densities of
10$^{6}$ cm$^{-3}$ and the field traced by line polarization at 10$^{2}$ -
10$^{3}$ cm$^{-3}$.} However, we were only able to compare our calculation with
polarization observations of CO molecular transitions. We believe that polarization
information from additional molecules may probe the magnetic field geometry 
at intermediate densities giving a more accurate picture of the field morphology
in star forming regions.

{\normalsize Having information about the magnetic field at different
densities is useful to test the gravitational state of equilibrium for the
cloud. Polarization mapping can allow the Chandrasekhar-Fermi
method to be used to obtain an estimate of the magnetic field strength and
test whether the envelope is magnetically supported or not. }The envelope
of DR21(OH) appears to be magnetically supported.

This research has been supported partially by NSF grants AST 02-05810 and
AST 99-88104.

\clearpage

{\normalsize \begin{deluxetable}{cccc}
\tablecolumns{4}
\tablewidth{0pc}
\tablenum{1}
\tablehead{
\colhead{Offsets in arcsec} & \colhead{$\phi_{J=2 \rightarrow1}$} &
\colhead{Offsets in arcsec} & \colhead{$\phi_{J=1 \rightarrow0}$}
}
\tablecaption{Position angle for CO $J=2 \rightarrow1$ and CO $J=1 \rightarrow
1$ at
velocity channel map $v=$ -10 km s$^{-1}$.
Data was interpolated at a tolerance of 0.2$^{\prime \prime}$
 that corresponds to approximate 2.7 $\times$
10$^{-4}$ pc using a distance to DR21(OH) of 3 kpc.}
\startdata
(1.0,2.0)  & 78$\pm$5 & (1.0,2.0)  & 3.6$\pm$9.2 \\
(-0.6,2.0) & 77$\pm$5 & (-0.6,2.0) & 4$\pm$9.6   \\
(4.2,4.0)  & 87$\pm$7 & (4.0,4.0)  &-0.2$\pm$8.9 \\
(2.6,4.0)  & 85$\pm$6 & (2.5,4.0)  &-5.2$\pm$6.2 \\
(1.0,4.0)  & 83$\pm$7 & (1.0,4.0)  &-6.1$\pm$5.3 \\
(-0.6,4.0) & 84$\pm$7 & (-0.5,4.0) &-4.6$\pm$6.2 \\
(9.0,6.0)  &-60$\pm$5 & (9.0,6.0)  &15.3$\pm$9.7 \\
(7.4,6.0)  &-70$\pm$6 & (7.5,6.0)  &13.0$\pm$7.6 \\
(5.8,6.0)  &-85$\pm$8 & (6.0,6.0)  & 6.7$\pm$6.9 \\
(4.2,6.0)  & 88$\pm$8 & (4.2,6.0)  &-3.5$\pm$5.5 \\
(2.6,6.0)  & 87$\pm$8 & (2.5,6.0)  &-9.3$\pm$4.1 \\
(9.0,8.0)  &-50$\pm$7 & (9.0,8.0)  &18.4$\pm$9.6 \\
(7.4,8.0)  &-62$\pm$8 & (7.5,8.0)  &14.1$\pm$7.6 \\
(5.8,8.0)  &-76$\pm$9 & (6.0,8.0)  & 8.0$\pm$6.6 \\
\enddata
\end{deluxetable}
}

\clearpage

{\normalsize \newpage}

\begin{figure}
\label{1} 
\figurenum{1} 
\includegraphics
[angle=-90,scale=0.6]{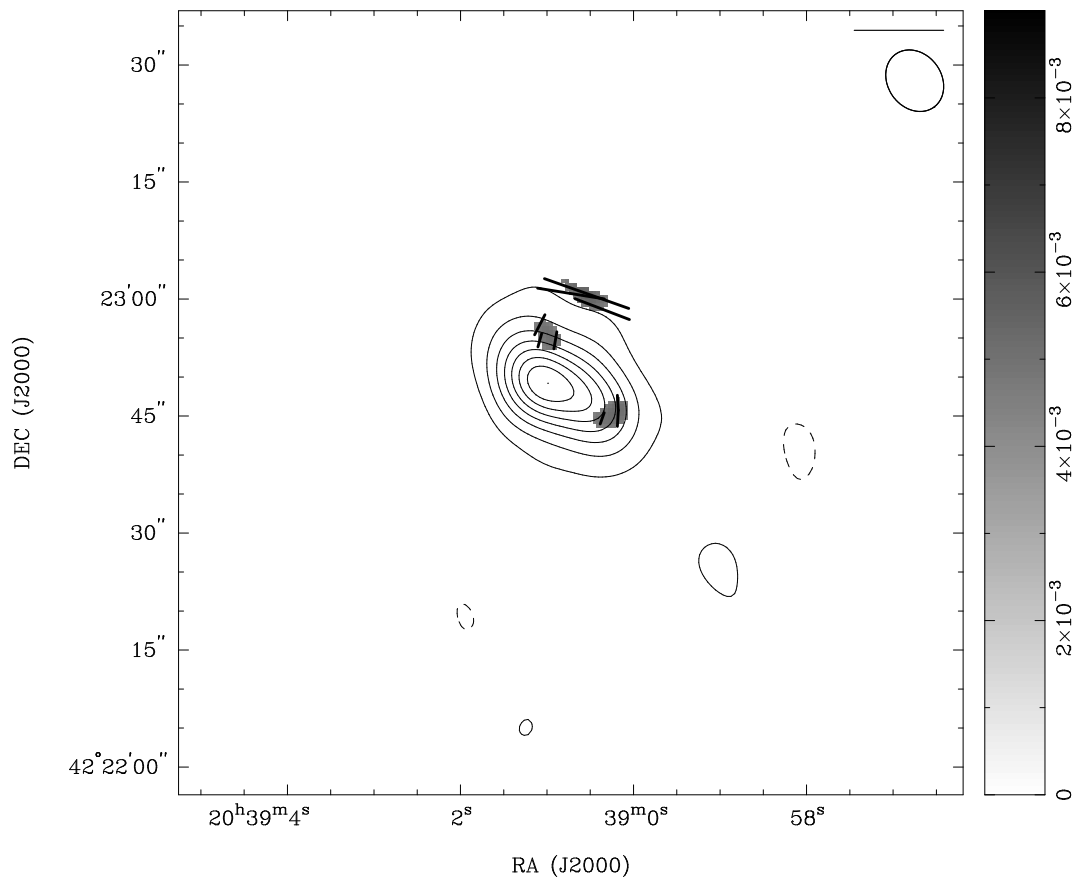} \caption{Polarization map of DR21(OH) at 3
mm. The contours represent Stokes I at -0.02, 0.02, 0.04, 0.07, 0.09, 0.11,
0.14, 0.16 and 0.18 Jy beam$^{-1}$. The pixel gray
scale show $3\sigma$ polarized intensity ($\sqrt{Q^{2}+U^{2}}$) 
for the dust continuum emission also
in Jy beam$^{-1}$,
while the black vectors are the dust fractional polarization. The length of the vectors
are proportional to fractional polarization with the length of the bar at the top-right 
corresponding to a fractional polarization of 0.33.}%
\end{figure}

\clearpage

\begin{figure}
\label{2} \figurenum{2} \includegraphics
[angle=-90,scale=0.6]{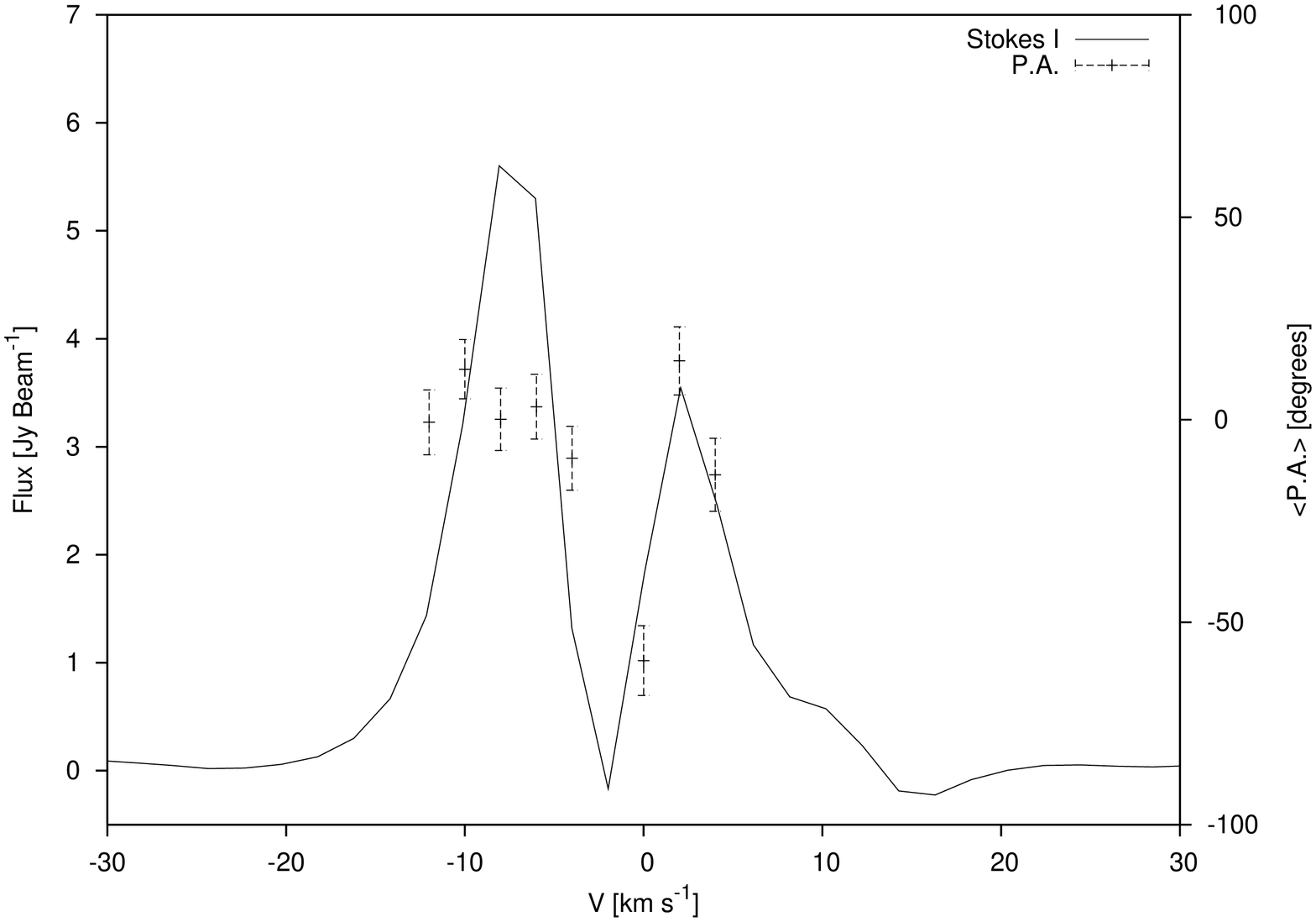} \includegraphics
[angle=-90,scale=0.6]{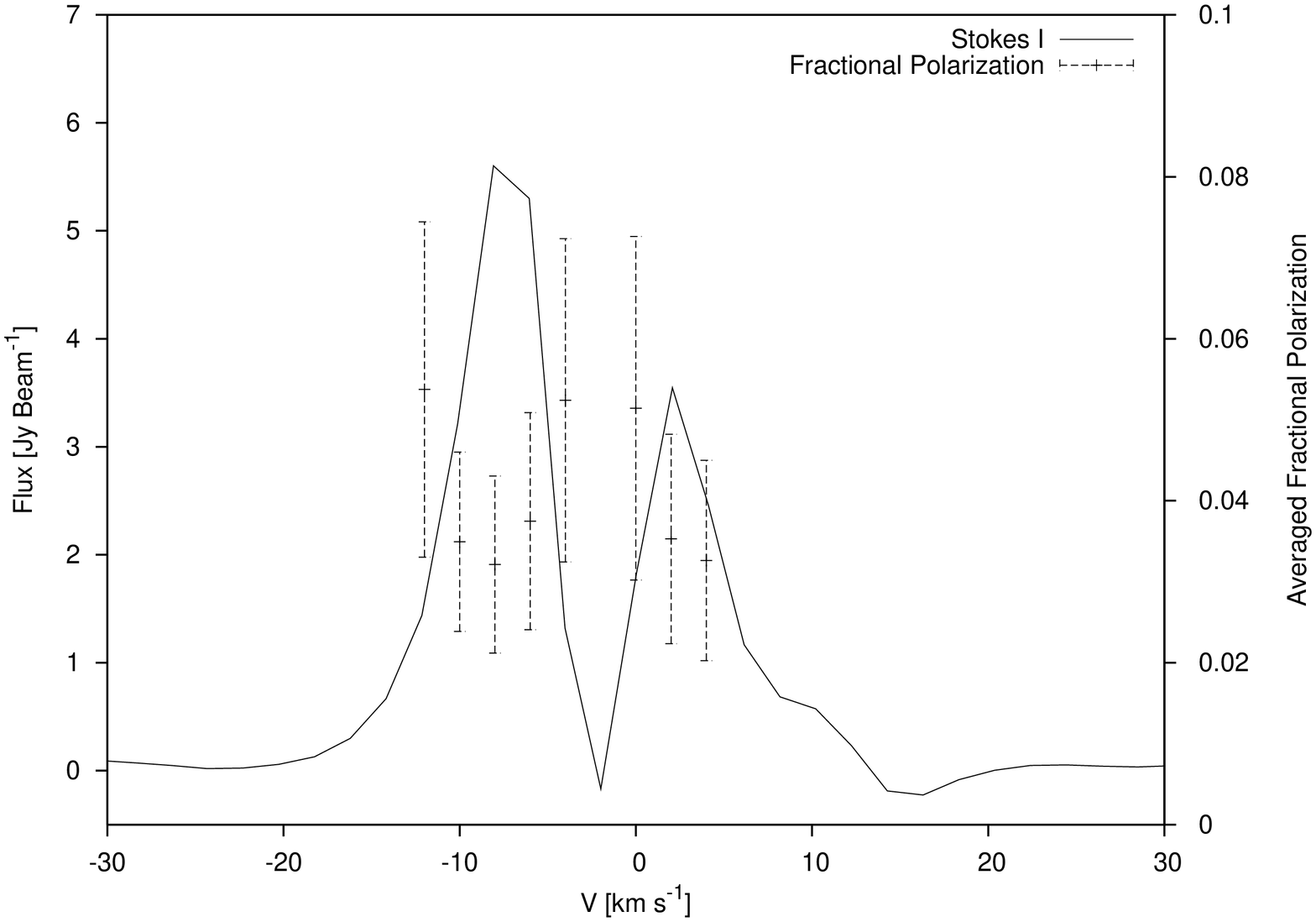} \caption{The spatially
averaged Stokes I spectrum from the CO $J=1 \rightarrow 0$ emission.
We observe a peak of 5.6 Jy beam$^{-1}$ 
at $v_{lsr}=-8$ km s$^{-1}$, while the minimum is located at
$v_{lsr}=-2$ km s$^{-1}$. ({\em upper panel}). Plotted over is the averaged P.A. over
the Stokes I spectrum. Most of the velocity channels have P.A.
that are orthogonal to the \citet{Lai2003} result. ({\em lower panel}) The integrated
fractional polarization is plotted over the Stokes I spectrum. }%
\end{figure}

\clearpage

\begin{figure}
\label{3} 
\figurenum{3} 
\includegraphics[angle=-90,scale=0.6]{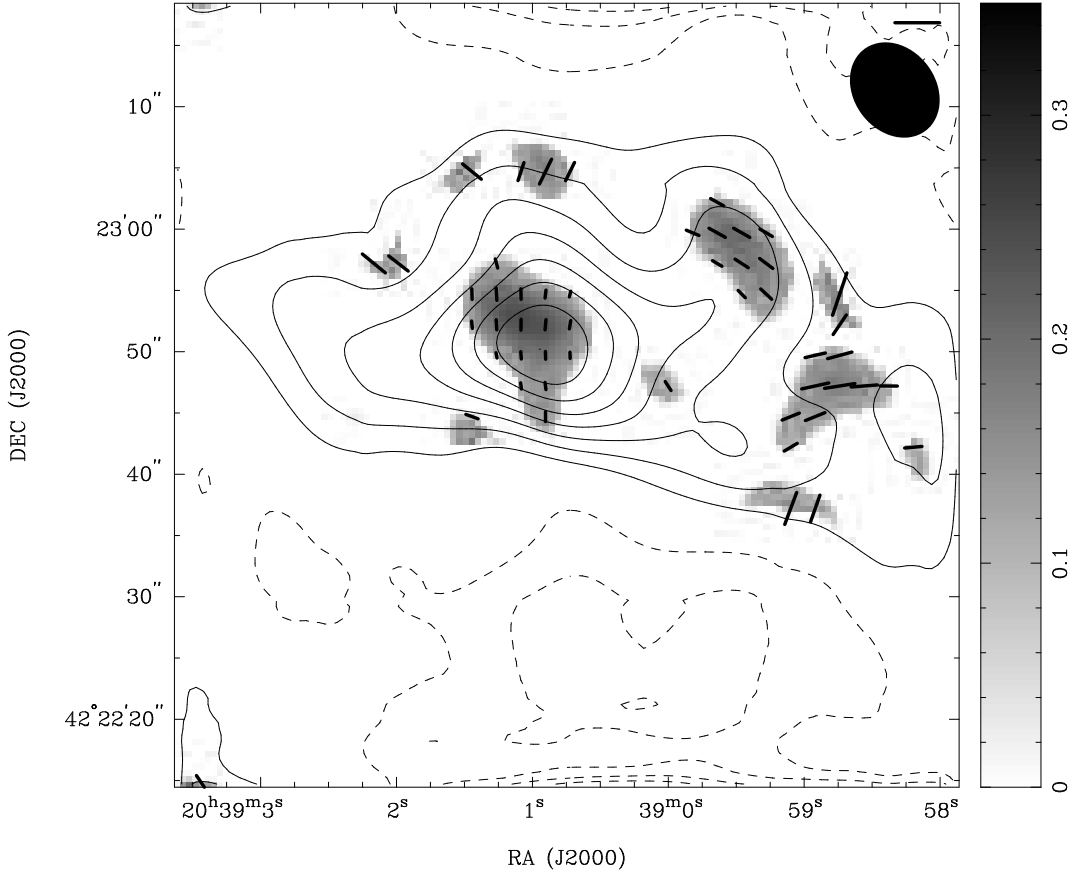} 
\caption{The CO
$J=1 \rightarrow0$ polarization emission from DR21(OH) at $v_{lsr}=$-8 km
s$^{-1}$. The contours represent Stokes I at -6.15, -4.61, -3.08, -1.54, 1.54, 3.08, 4.61,
6.15, 7.7, 9.23, 10.77 and 12.31 Jy beam$^{-1}$. The gray scale represents
$3\sigma$ polarized emission ($\sqrt{Q^{2} + U^{2}}$) also in Jy beam$^{-1}$. 
The vectors are a representation of the fractional polarization, the length of the bar
at the top-right corresponds to a fractional polarization of 0.086.
 The most widespread
polarization pattern occurs at this velocity, which coincides with the peak of
the CO emission (Stokes I image). }
\end{figure}

\clearpage

\begin{figure}
\label{4} 
\figurenum{4} 
\includegraphics[angle=-90,scale=0.6]{f5.eps} 
\epsscale{0.4} 
\caption{The
CO $J=1 \rightarrow0$ polarization emission from DR21(OH) at $v_{lsr}=$-10 km
s$^{-1}$. The contours represent Stokes I at -2.73, -1.82, -0.91, 0.91, 1.82, 
2.73, 3.64, 4.56, 5.47, 6.38 and  7.3 Jy beam$^{-1}$. The gray scale represent
$3\sigma$ polarized emission ($\sqrt{Q^{2} + U^{2}}$) also in Jy beam$^{-1}$. 
The vectors are a representation of the fractional polarization,
 the length of the bar at the top-right corresponds to a
fractional polarization of 0.064.
This  map represent the same
velocity channel presented by \citet{Lai2003}. Comparing with \citet{Lai2003}
at the same velocity channel we see that most of the polarization is
perpendicular to the CO $J=2 \rightarrow1$ polarization.}%
\end{figure}

\clearpage

\begin{figure}
\label{5} 
\figurenum{5} 
\includegraphics[angle=-90,scale=0.6]{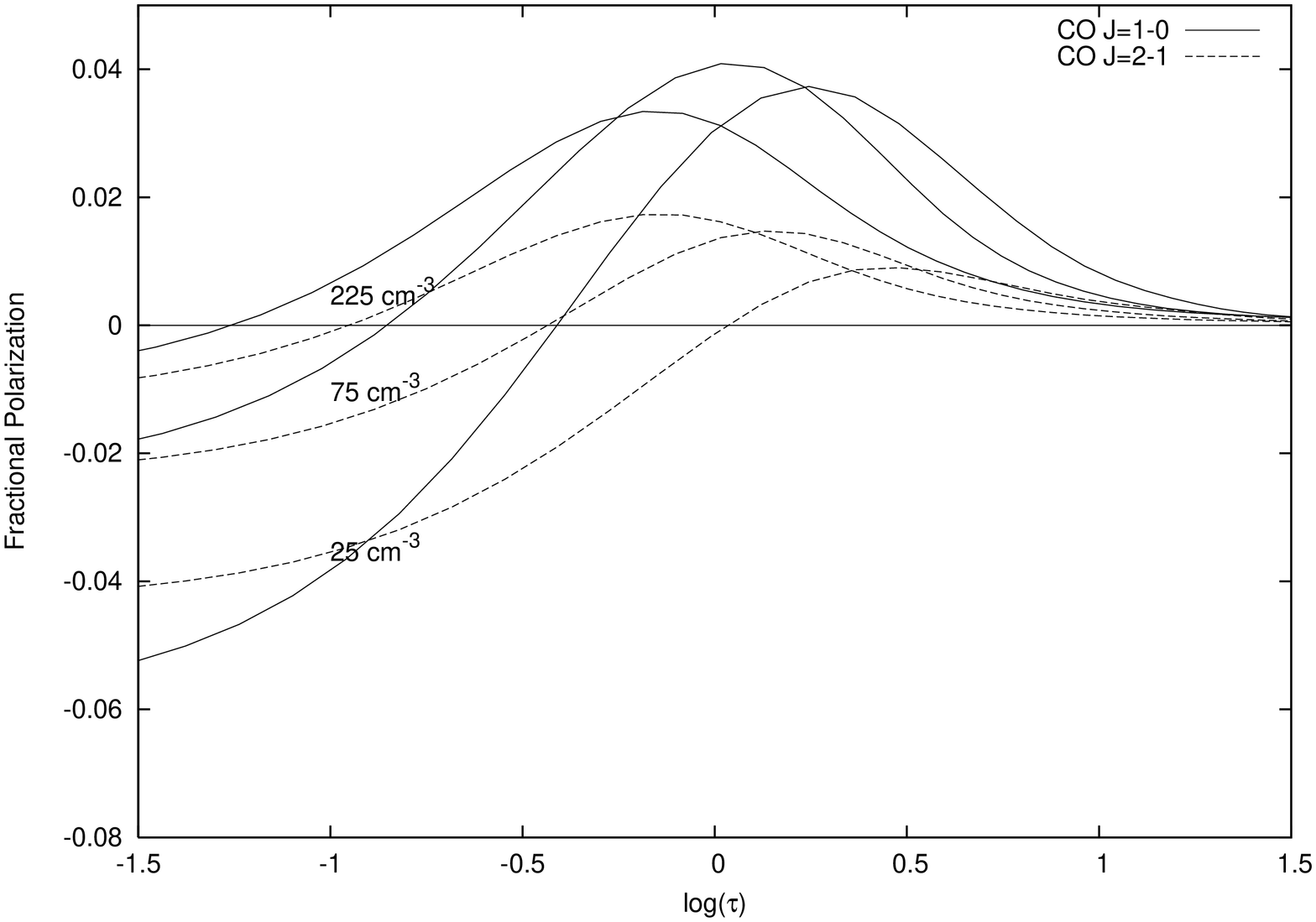}
\includegraphics[angle=-90,scale=0.6]{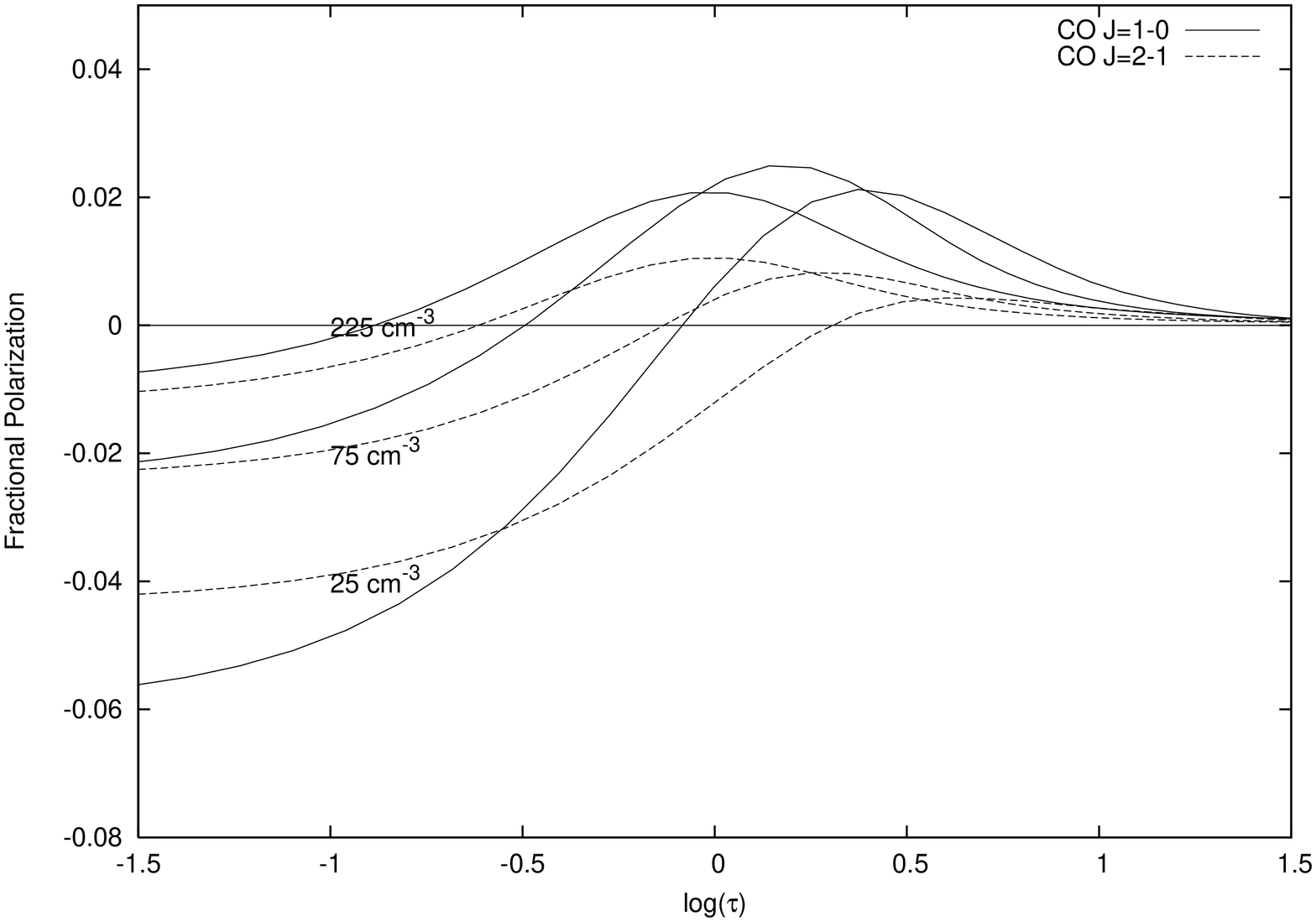}
\caption{Fractional
polarization $P$ calculated for the CO transitions $J=1\rightarrow0$ and $J=2\rightarrow
1$ as a function of the LVG optical depth for the CO $J=1\rightarrow0$ transition.
Here,
$T_{gas}=$28 K and $T_{source}=$42 K. Each pair of curves represents 
calculations for a different
density of molecular hydrogen---25 cm$^{-3}$, 75 cm$^{-3}$ and 225 cm$^{-3}$.
In the {\em upper panel}, the velocity gradient is characterized by $\alpha$=0.1.
In order to test the robustness
of the calculation, we decreased the degree of anisotropy by setting $\alpha$=0.3.
This is shown in the {\em lower panel}.} 
\end{figure}

\clearpage

\begin{figure}
\label{6} 
\figurenum{6} 
\includegraphics[angle=-90,scale=0.6]{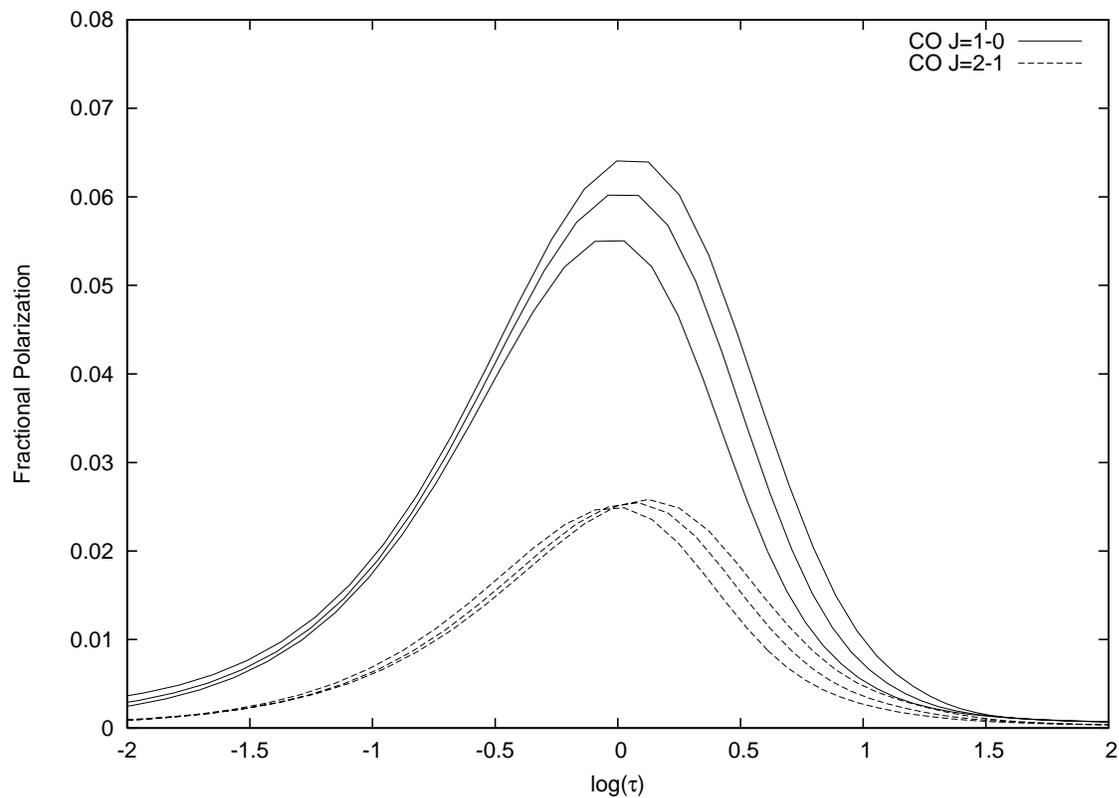}
\caption{Fractional polarizations $P$ calculated  
for the CO transitions $J=1\rightarrow0$ and $J=2\rightarrow
1$ are shown  as a function of   
$\tau$ for the molecular hydrogen densities 25, 75 and 225 cm$^{-3}$
when there is no compact, external source of continuum radiation.  
Here, the direction of the polarizations is determined by the
direction of the velocity gradient and is perpendicular to the
direction of the magnetic field. Note that the polarizations of 
the CO $J=1\rightarrow0$ and $J=2\rightarrow1$  transitions 
are never orthogonal (i. e. they always have the same sign.)}%
\end{figure}

\clearpage

\begin{figure}
\label{7} 
\figurenum{7} 
\includegraphics[angle=-90,scale=0.6]{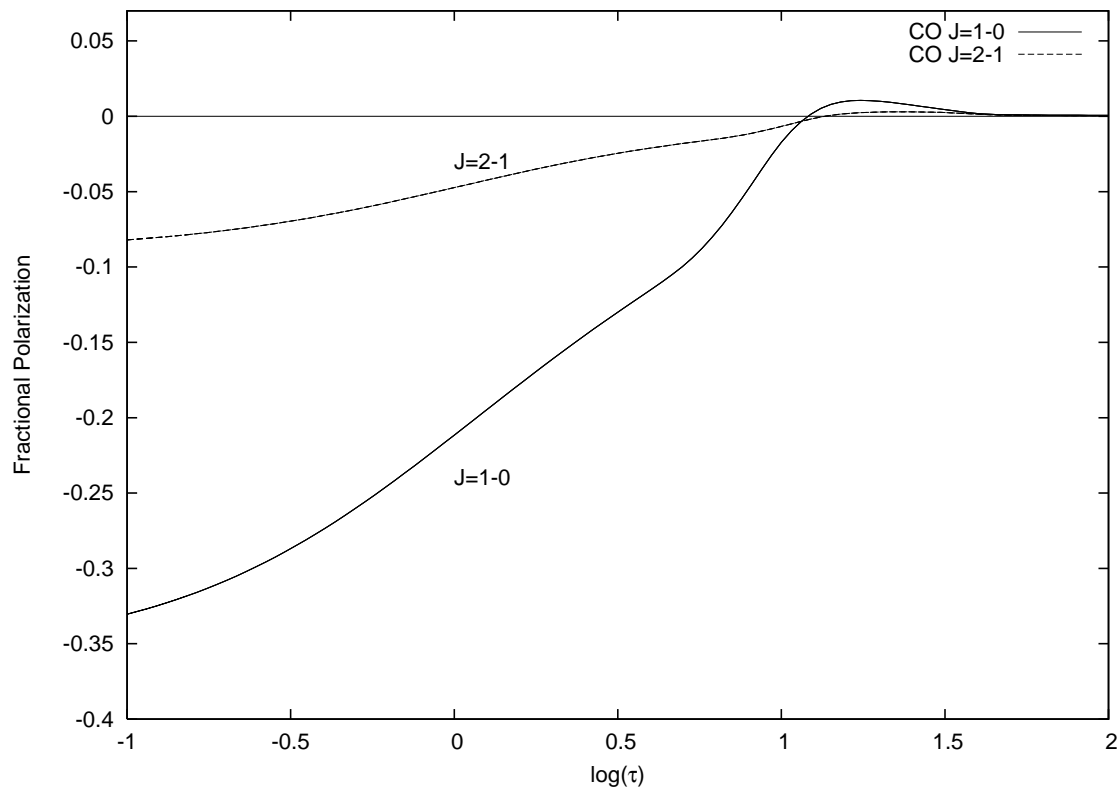} 
\caption{Fractional
polarization $P$ for the CO transitions $J=1\rightarrow0$ and $J=2\rightarrow
1$ are shown as a function of $\tau$ 
when the collision rates used in Figure 5 are reduced by multiplying
by a factor 10$^{-5}$. For the relevant hydrogen densities (25 - 225 cm$^{-3}$),
the external continuum source dominates
in determining the direction of the polarization for both transitions
and the polarizations do not depend on density.
For both spectral lines, the
polarization is parallel to the magnetic field (negative sign) as expected, and converges 
to zero for large values of $\tau$.}%
\end{figure}

\end{document}